\documentclass[aps,prb,superscriptaddress,twocolumn,preprintnumbers,amsmath,amssymb]{revtex4}
\usepackage{color}
\usepackage{array}
\usepackage{amsmath}
\usepackage{amssymb}
\usepackage{epsfig}
\usepackage{graphicx}
\usepackage{bbm}

\newcommand{\ie}{{\emph{i.e.~}}}
\makeatletter

\newcommand{\Rmnum}[1]{\expandafter\@slowromancap\romannumeral #1@}
\makeatother
\newcommand{\imth}{\hspace{1pt}\mathrm{i}\hspace{1pt}}

\newcommand{\eg}{{\emph{e.g.~}}}

\newcommand{\tk}{\textbf{k}}
\newcommand{\tA}{\textbf{A}}

\begin{document}
\title{Quantum Hall effects in a Weyl Semi-Metal: possible application in pyrochlore Iridates}
\author{Kai-Yu Yang}
\author{Yuan-Ming Lu}
\author{Ying Ran}
\affiliation{Department of Physics, Boston College, Chestnut Hill, MA 02467}
\date{\today}

\begin{abstract}
There have been lots of interest in pyrochlore Iridates A$_2$Ir$_2$O$_7$ where both strong spin-orbital coupling and strong correlation are present. A recent LDA calculation\cite{PhysRevB.83.205101} suggests that the system is likely in a novel three dimensional topological semi-metallic phase: a Weyl semi-metal. Such a system has zero carrier density and arrives at the quantum limit even in a weak magnetic field. In this paper we discuss two novel quantum effects of this system in a magnetic field: a pressure-induced anomalous Hall effect and a magnetic field induced charge density wave at the pinned wavevector connecting Weyl nodes with opposite chiralities. A general formula of the anomalous hall coefficients in a Weyl semi-metal is also given. Both proposed effects can be probed by experiments in the near future, and can be used to detect the Weyl semi-metal phase.
\end{abstract}

\maketitle
\section{Introduction}
Experimental realizations of two-dimensional massless Dirac electrons in condensed matter systems have generated a lot of interest. These include the intrinsic two-dimensional graphene system\cite{novoselov-2005-438}, as well as the surface of a three-dimensional topological insulator\cite{Moore_Review_2010,RevModPhys.82.3045,qi:33}. One of the many exciting phenomena for these systems is their anomalous response to an external magnetic field. For example, the room-temperature integer quantum hall effect\cite{Zhang2005,Novoselov02152007} has been observed in graphene system. 

A minimal model for a two dimensional Dirac electronic system is $H=v_F(p_x\sigma_x+p_y\sigma_y)$, where $\vec p$ is momentum and $\vec \sigma$ are Pauli matrices. Clearly a mass term $m\sigma_z$ will generate an energy gap for the electronic structure. One can ask whether mass terms will appear in the experimental systems mentioned above, in which case the linear dispersive band touching points, the Dirac nodes, will be destroyed. In these systems, it turns out that the Dirac nodes are protected by extra physical symmetries apart from the lattice translational symmetry. For example, in the case of the surface states of a topological insulator, it is protected by time-reversal symmetry.

Recently a remarkable theoretical work\cite{PhysRevB.83.205101} indicates that, a novel \emph{three-dimensional} relativistic electronic structure, the Weyl semi-metal phase, is likely to be realized in pyrochlore Iridates A$_2$Ir$_2$O$_7$ where A=Yttrium, or a Lanthanide element. On one hand, similar to graphene, the electronic dispersion of a Weyl semi-metal is characterized by a set of linear-dispersive band-touching points of \emph{two} adjacent bands, the Weyl nodes. On the other hand, there are important differences between the 3D Weyl nodes and the 2D Dirac nodes in graphene, because the Weyl nodes are protected by the topology of the band structure. One direct way to see this is to write down the effective hamiltonian in the neighborhood of a Weyl node: $H=v_F(p_x\sigma_x+p_y\sigma_y+p_z\sigma_z)$. The three Pauli matrices are used up and there is simply no local mass term. Consequently, as long as there is no translational-symmetry-breaking inter-valley mixings between different Weyl nodes, the Weyl semi-metal phase is robust for arbitrary perturbation.

The concept of Weyl fermions was firstly introduced in high energy physics and has been used to describe neutrinos. The possible realizations of Weyl electronic structures in condensed matter systems and their superconducting analogs were discussed by various authors\cite{Nielsen1983389,PhysRevB.58.2788,volovik:book}. In fact, the original attempt to realize Weyl fermions in 3D lattice systems results in the famous fermion-doubling theorem, which dictates the total number of Weyl nodes must be even\cite{Nielsen198120}. This is related to another famous phenomena, the Adler-Bell-Jackiw anomaly (or chiral anomaly)\cite{Nielsen1983389}. Weyl Fermions have its handiness or chirality. Chiral anomaly states that a quantized space-time electromagnetic field event would pump quantized electric charge from a node with positive chirality to one with negative chirality. Thus the number nodes of positive chirality must equal those with negative chirality; totally one must have even number of nodes.

Because the Dirac spectrum is known to have anomalous response to a magnetic field, a natural question to ask is: what is the response of a Weyl semi-metal in a magnetic field? Motivated by the fact that Weyl semi-metal is a novel phase of matter whose experimental signatures are of fundamental interest, and also by the recent experiment efforts on the pyrochlore Iridates, we study the effects of an external magnetic field on a Weyl semi-metal.

Let us state the main results of this work. We find two novel quantum effects of a Weyl semi-metal in a magnetic field: a pressure-induced anomalous Hall effect and a magnetic field induced charge density wave at the pinned wavevector that connects nodes with opposite chiralities. A general formula of the anomalous hall conductivity in a Weyl semi-metal is also given. We also apply these results to the proposed Weyl phase in pyrochlore Iridates and address the experimental relevant questions in these specific systems.

Pyrochlore Iridates A$_2$Ir$_2$O$_7$ have attracted a lot of attentions both experimentally and theoretically\cite{JPSJ.70.2880,0953-8984-13-23-312,JPSJ.76.043706,PhysRevB.83.180402,2010NatPh...6..376P,PhysRevB.83.205101,PhysRevB.82.085111,PhysRevB.83.165112}. Because of the feature of the Ir$^{4+}$ ion, these systems are in a novel regime where strong spin-orbital coupling, strong correlation as well as geometric frustration are present, and new physics may emerge. As temperature is lowered, the A=Eu, Sm, Nd systems experience metal-insulator phase transitions\footnote{The resistivities of these ``insulators'' increase steeply as temperature is lowered. But the absolute values of the resistivities in the low temperature limit are not small at all $\rho\sim 10^{0\sim 1}\Omega$cm, and far from a typical insulator.} that are clearly associated with a singularity in the magnetic susceptibility, suggesting magnetic ordering\cite{JPSJ.70.2880,0953-8984-13-23-312,JPSJ.76.043706}. Recent $\mu$SR measurement on A=Eu system, where metal-insulator transition occurs at $~120$K, suggests large static ordering moment $\sim 1\mu_B$ from Ir$^{4+}$\cite{PhysRevB.83.180402}. Because of the lack of neutron scattering data, the magnetic structure of the low temperature phases remains unclear. 

On the theoretical side, a calculation based on a microscopic model suggests that the insulating phase can be a novel spin liquid without magnetic ordering\cite{2010NatPh...6..376P}. A more recent LDA+U calculation, however, shows that, depending on strength of correlation, the system can be in a novel 3D semi-metal phase associated with a ``4-in, 4-out'' anti-ferromagnetic order\cite{PhysRevB.83.205101}. If the proposed Weyl semi-metal phase is realized in stoichiometric clean pyrochlore Iridates, chemical potential will be automatically tuned to the Weyl nodes. One clear prediction was made in Ref\cite{PhysRevB.83.205101} where authors show that there are topologically protected surface chiral Fermi-arc, which can be detected in ARPES (angle-resolved photoemission spectroscopy) experiments. However, it is unclear whether this 3D material is experimentally friendly in terms of surface sensitive probes. Part of the goals of this paper is to find the bulk signatures of the Weyl phase.

\section{Pressure-induced anomalous Hall effect}
\begin{figure}
 \includegraphics[width=0.45\textwidth]{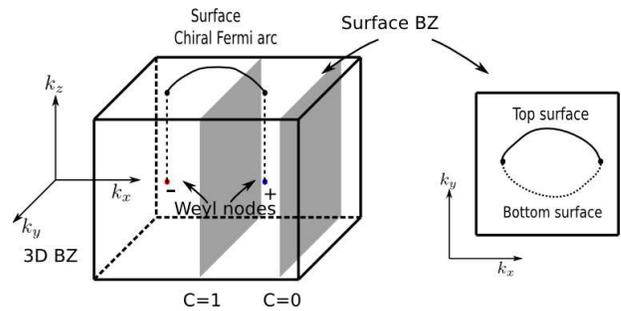}
\caption{(color online) Weyl nodes in the two-band model Eq.\ref{eq:two-orbit}. The Chern number for the 2D band structure $C$ at a given $k_x$ is jumping by 1 across the nodes. As a result, there are surface chiral Fermi arcs. The arcs on the top and bottom surfaces form a closed 2D Fermi surface.}
\label{fig:semi-classical}
\end{figure}

One way to view the Weyl node is that it is a monopole of the Berry-curvature\cite{Fang03102003,PhysRevLett.93.206602,PhysRevB.83.205101}. For example, let us consider a simple half-filled 3D two-band model:
\begin{align}
H_{\boldsymbol{k}} = & [2t_{x}(\cos{k}_{x} - \cos{k_{0}}) + m(2-\cos{k_{y}}-\cos{k_{z}})]\sigma_{x} \nonumber \\
 &+ 2t_{y}\sin{k_{y}}\sigma_{y} +2t_{z} \sin{k_{z}} \sigma_{z},
\label{eq:two-orbit}
\end{align}
where $\sigma$ is the spin of the electron. This model breaks time-reversal symmetry and hosts two Weyl nodes in the bulk Brillouin Zone(BZ) at $\vec{\mathbf{P}}=\pm(k_0,0,0)$, related by inversion symmetry (see Fig.\ref{fig:semi-classical}). If we fix $k_x$, $H_{k_x}(k_y,k_z)$ can be viewed as a 2D band structure, which is fully gapped when $k_x\neq \pm k_0$ and its Chern number $C_{k_x}$, or TKNN index\cite{PhysRevLett.49.405} is well-defined. It is easy to show that $C_{k_x}=1$ when $k_x\in(-k_0,k_0)$ and $C_{k_x}=0$ otherwise. In this sense the Weyl nodes can be viewed as integer quantum hall plateau transition as $k_x$ is tuned. Because $C_{k_x}$ is an integration of the Berry's curvature, the jump of $C_{k_x}$ at a Weyl node dictates that it is a magnetic monopole of the Berry's curvature, positively(negatively) charged if its chirality, defined as the handiness of the momentum axis in front of the $\sigma_{x,y,z}$ matrices, is positive(negative). A direct consequence of these monopoles is that, on the surface not perpendicular to $k_{x}$ direction, for instance, the $x-y$ surface, there must be a chiral Fermi surface connecting the Weyl nodes in the surface BZ\cite{PhysRevB.83.205101} -- a Fermi ``arc''.

The association of a Weyl node with the jump of the Chern number indicates that the system may have a large anomalous hall effect. (Anomalous hall effect associated with monopoles in momentum space of ferromagnetic systems was discussed by Fang et al\cite{Fang03102003}.) Indeed in the two-band model Eq.(\ref{eq:two-orbit}), from the existence of the surface modes, which correspond to one edge state for every $2\pi/(2k_0)$ $y-z$ layers, anomalous hall effect occurs with $\sigma_{yz}=\frac{e^2}{h}2 k_0$. When the two nodes are moved to the BZ boundary and annihilated, the system becomes an quantized 3D anomalous hall state.

\begin{figure}
 \includegraphics[width=0.3\textwidth]{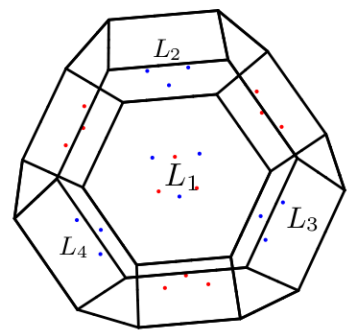}
\caption{(color online) The nodes of the proposed Weyl metal phase\cite{PhysRevB.83.205101} of A$_2$Ir$_2$O$_7$ in its first BZ. Red dots and blue dots are of opposite chirality. The direction out or paper plane is along [1,1,1] direction.}
\label{fig:BZ}
\end{figure}

In the pyrochlore Iridates A$_2$Ir$_2$O$_7$, however, the proposed Weyl phase hosts 24 nodes\cite{PhysRevB.83.205101} (see Fig.\ref{fig:BZ}), all related by the lattice cubic symmetry. Are there anomalous hall effects in this phase?

In a general 3D crystal, anomalous hall effect is characterized by a momentum space vector\cite{Halperin_3D_Quantum_Hall} $\vec \nu$, the Chern vector. The anomalous hall conductivity is given by $\vec \nu$ via: $\sigma_{ij}=\frac{e^2}{2\pi h}\epsilon_{ijk}\nu_k $. Haldane\cite{PhysRevLett.93.206602} shows that anomalous hall conductivity of the ground state of a 3D electronic structure can be expressed as an integration of the Berry's curvature of the filled electronic states:
\begin{align}
 \sigma_{ij}=\frac{e^2}{\hbar}\frac{1}{\Omega N}\sum_{\vec k,a}\mathcal{F}_{ij}^a n_{a}(\vec k,\mu),\label{eq:haldane}
\end{align}
where $N$ is the number of unit cells, each of which has volume $\Omega$. $\mathcal{F}_{ij}^a$ is the well-known $U(1)$ Berry's curvature in band $a$: $F_{ij}^a=\partial_iA_j^a-\partial_jA_i^a$ where $A^a_i=-i\langle u^a_{\vec k}|\partial_{k_i}|u^a_{\vec k}\rangle$ and $|u^a_{\vec k}\rangle$ is the Bloch state. This means $\vec\nu$ is completely determined by the band structure and Fermi level. If the 3D system is fully gapped, one can show that $\vec\nu$ must be a reciprocal lattice vector. In this case, let the quantized $\vec \nu=\vec G$; and the system can be viewed as a stacking of 2D quantized anomalous hall layers along the $\vec G$ direction.

Here we provide a remarkably simple formula of the anomalous Hall coefficient in a general Weyl semi-metal:
\begin{align}
 \vec \nu_{node}=\sum_i (-)^{\xi_i} \vec{\mathbf{P}_i}.\label{eq:nu}
\end{align}
Here $\vec \nu_{node}$ is the Chern vector of the ground state of a Weyl semi-metal, where the chemical potential is at the nodes. $i$ labels all different nodes, $\vec{\mathbf{P}_i}$ are their momentum, and $\xi_i$ are their chiralities. Note that here we do not restrict $\vec{\mathbf{P}_i}$ to be in the first Brillouin Zone. As a result Eq.(\ref{eq:nu}) cannot be used to determine $\vec \nu$ completely; instead it only determines the fractional part of $\vec \nu$ unambiguously because one can always add a fully filled gapped band with Chern number.

\begin{figure}
 \includegraphics[width=0.45\textwidth]{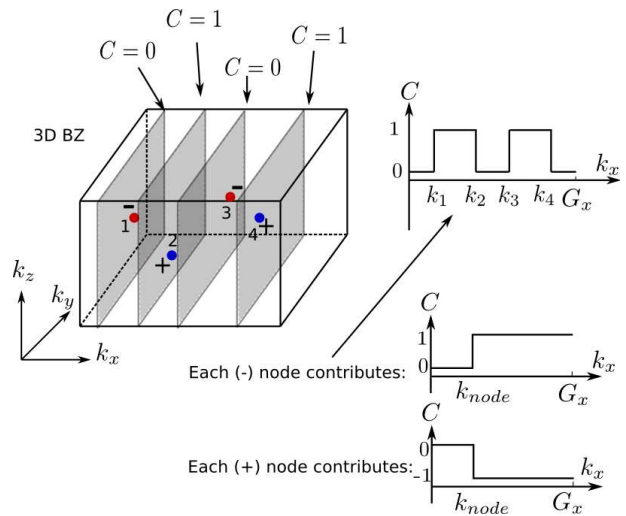}
\caption{(color online) Schematic illustration of the proof of the general formula Eq.(\ref{eq:nu}).}
\label{fig:AHE_formula}
\end{figure}

The proof of this formula is quite straightforward starting from Eq.(\ref{eq:haldane}). For simplicity, let us assume that there are four Weyl nodes located in the 3D BZ, as shown in Fig.\ref{fig:AHE_formula}. Let us study $\nu_x$ first. Similar to what we mentioned in the two-band model, we cut the 3D BZ into 2D slices for various value of $k_x$. Unless the cut goes through the nodes, the 2D band structure $H_{k_x}(k_y,k_z)$ is fully gapped, and the Chern number $C_{k_x}$ is well defined. The total $\nu_x$ should be integration $\int_0^{G_x} C_{k_x}dk_x$. Because the Weyl node is a monopole (anti-monopole), one easily convinces oneself that every node at $\vec k_{node}$ with negative(positive) chirality contributes $C_{k_x}=\Theta(k_x-k_{node,x})$ ($C_{k_x}=-\Theta(k_x-k_{node,x})$), where $\Theta$ is the step function. After integration one proves Eq.(\ref{eq:nu}) for $x$-direction, and similarly for $y,z$-directions.

Plugging in all the 24 nodes' momenta and chiralities for the proposed Weyl phase in pyrochlore Iridates, Eq.(\ref{eq:nu}) gives vanishing anomalous hall effect $\vec \nu=0$. There is no surprise here because of the cubic symmetry of the system. $\vec\nu$ must vanish because it cannot choose a special direction in momentum space.

What if the lattice symmetry is not cubic? This can be realized, for example, by applying a uni-axial pressure along the [1,1,1] direction. In this case the [1,1,1] direction is special and symmetry consideration allows nonzero $\vec\nu\parallel[1,1,1]$. In the following we show that this indeed happens with $|\vec\nu|$ as a linear function of the pressure enhancement $P$ in the low $P$ limit. We predict that a pressure $\lesssim 1$GPa, which typically modifies the electronic hopping integrals in the band structure by a few percent, can induce a large anomalous hall effect, corresponding to a few percent of integer quantum hall conductance per atomic layer. \emph{This pressure-induced large anomalous hall effect with its linear $P$ dependence is an intrinsic signature a Weyl semi-metal phase} when the original crystal symmetry dictates zero anomalous hall effect, and can be used to detect it in experiment.

The cubic symmetry of A$_2$Ir$_2$O$_7$ is broken to trigonal symmetry by a pressure along [1,1,1] direction. As a result the 24 nodes are no longer all related by symmetry. The nodes will shift in momentum space and chemical potential $\mu$ will no longer be at the node (referred to as self-doping from now on). Because the summation of all filled states in Eq.(\ref{eq:haldane}) can be separated into the summation of all states below the Weyl nodes, and the summation due to self-doping, the change of the Chern vector under a pressure have two contributions $\delta \vec \nu=\delta\vec \nu_{node}+\delta\vec \nu_{doping}$, where $\delta\vec \nu_{node}$ is due to the shift of nodes, and $\delta\vec \nu_{doping}$ is due to self-doping of the nodes. We will show $\delta\vec \nu_{node}\propto P$ and $\delta\vec \nu_{doping}\propto P^2$ in the low $ P$ limit with $\mu \propto P$. As a result, in the low $P$ limit, $\delta\vec \nu_{node}$ dominates and $\delta \vec \nu\propto P$.

We first discuss $\delta\vec \nu_{node}$. Because the proof of Eq.(\ref{eq:nu}) still goes through for the contribution of $\vec\nu$ from all the states below the nodes, it can be used to compute $\delta\vec \nu_{node}$.
It is then clear that $\delta\vec \nu_{node}\propto P$ because the shifts of the nodes generically will be a linear function of $P$. 

\begin{figure}
\centerline
{
\includegraphics[width = 0.45\textwidth]
{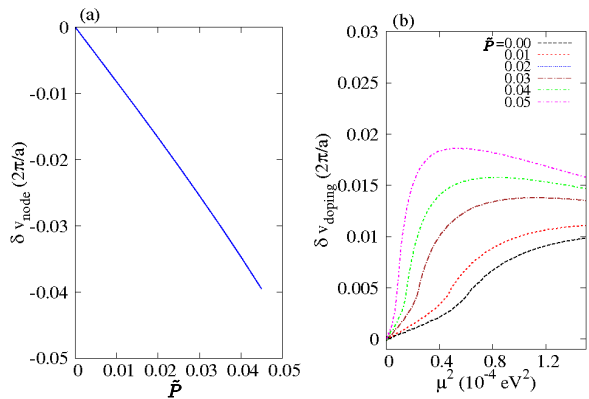}
}
 \caption
 {(color online)
Left panel: The $\tilde P$ modeling pressure along $\Gamma-L_{1}$ ([1,1,1]) direction vs Chern vector $\vec \nu$ contributed from the Dirac points shifts. ($ \vec \nu\parallel [1,1,1]$, the projection of $\vec \nu$ along $[1,1,1]$ is shown.)
Right panel: The $\vec \nu$ contributed from the self-doping happened on Dirac nodes around $L_{2,3,4}$ points.
}
\label{fig: pocket_mu_Iridate_tot}
\end{figure}

To confirm this claim, we have modeled the effect of pressure in the Weyl phase of A$_2$Ir$_2$O$_7$ by multiplying the hopping integrals along [1,1,1] direction by a factor $1+\tilde P$ ($\tilde P>0$) in the low energy $k\cdot p$ theory described in Ref\cite{PhysRevB.83.205101}:
 \begin{align}
H({\boldsymbol{q}},L_{i}) =& (\Delta + \frac{q_{z,i}^{2}}{2m_{1}} - \frac{q_{\perp,i}^{2} }{2m_{2}}) \sigma_{z} \nonumber \\
&+ (\beta q_{z,i} + \lambda q_{\perp,i}^{3} \cos{3 \theta}) \sigma_{x} + \lambda q_{\perp,i}^{3} \sin{3 \theta}\sigma_{y}
\label{eq:HL}
\end{align}
where $q_{z,i}, q_{\perp,i}$ is defined locally around each $L$ points with $q_{z,i}$ along the $\Gamma-L_{i}$ direction.
The three pair of Dirac points around $L$ points located at $q_{\perp}^{2} \sim 2m_{2}\Delta, q_{z,i} \sim \mp q_{\perp}^{3} \lambda / \beta$. By choosing
$\Delta = 0.18 \rm eV$, $\beta = 0.5 \rm eV$, $m_{1} = 0.5 \rm eV^{-1}$, $m_{2} = 0.5 \rm eV^{-1}$, $\lambda =1 eV$, $q$ dimensionless within $(-\pi,\pi]^{3}$, and appropriate $\theta$ orientation, this hamiltonian roughly captures the locations and energy scales of the Weyl nodes. To simulate the effect of pressure, in Eq.\ref{eq:HL} we multiply each term having $q_{z,1}$ (not $q_{z,i}$) by a factor of $1+\tilde P$ corresponding to increase of hopping integral along $\Gamma-L_{1}$ direction. $\delta\vec \nu_{node}$ is computed by Eq.(\ref{eq:nu}). As shown in Fig~\ref{fig: pocket_mu_Iridate_tot} (a), even a small $\tilde P$ leads to a substantial change of $\delta\vec\nu_{node}\propto \tilde P\propto P$. 

Next we consider $\delta\vec \nu_{doping}$. Naively this would also be a linear function of $P$, presuming for a given node $\delta\vec \nu_{doping}\sim \int_0^{k_F} \vec B(\vec k) k^2dk\sim k_F$, where $\vec k$ is momentum measured from the node, $B_i(\vec k)\equiv\frac{1}{2}\epsilon_{ijk}\mathcal{F}_{jk}$ and $|\vec B(\vec k)|\sim \frac{1}{k^2}$ since the node is a magnetic monopole. Generically $k_F$ would be a linear function of $P$. However, a closer look shows that this linear term actually vanishes. The simplest way to see this is to consider the low energy effective theory of a Weyl node: $H=\sum_i(\vec v_i\cdot \vec k)\sigma_i-\mu$. One can introduce a formal ``time-reversal'' anti-unitary transformation which sends $\vec k\rightarrow -\vec k$, and also flips all the signs of the Pauli matrices. This leaves $H$ invariant. Based on ``time-reversal'' symmetry, it is easy to show that $\vec B(-\vec k)=-\vec B(\vec k)$ and thus the linear $k_F$ term vanishes. A high order term in dispersion introduced by breaking this symmetry will generally lead to a nonzero $\delta\vec \nu_{doping}$.

Therefore we proved that $\delta\vec \nu_{doping}$ is completely due to the deviation from the Dirac dispersion and $\delta\vec \nu_{doping}\propto P^2$ at the leading order. To confirm this claim, we have also modeled self-doping in the $k\cdot p$ theory of A$_2$Ir$_2$O$_7$. After pressure in applied, the 24 nodes are split into three clusters 6(close to $\vec L_1$)+(6+12) (close to $\vec L_{2,3,4}$). The nodes within each cluster are related by trigonal symmetry. Because $k\cdot p$ theory Eq.(\ref{eq:HL}) only describes physics around each $\vec L_i$ point, there will be two undetermined relative chemical potentials between the three clusters of nodes. For simplicity, we choose the cluster of 6 nodes(close to $\vec L_1$) to be undoped, and the chemical potential of the cluster of 12 nodes to be $\mu$, while the chemical potential of other cluster of 6 nodes is determined by charge neutrality. We plot $\delta\vec \nu_{doping}$ for various value of $\tilde P$ in Fig.~\ref{fig: pocket_mu_Iridate_tot}(b) and it is clear that $\delta\vec \nu_{doping}\propto \mu^2$ and thus $\propto P ^2$ in the low $P$ limit. When $\mu$ is large $\delta\vec \nu_{doping}$ is controlled by the non-universal high energy band structure. 

We can estimate the magnitude of the pressure-induced anomalous hall conductivity. $1\%$ change of the hopping along [1,1,1] direction ($\tilde P=0.01$) induces $\nu\sim 0.01(2\pi)/a$, namely $\sigma_{AH}\sim 4(\Omega^{-1}\mbox{cm}^{-1})$. 
However if the uniaxial pressure is applied along [1,0,0] direction, the cubic symmetry is broken down to tetragonal symmetry and $ \vec \nu$ remains zero due to symmetry. The anomalous hall effect induced by pressure along [1,1,1]-direction is a rather stable (w.r.t disorders and temperature) signature of the proposed Weyl semi-metal phase, and can be used to detect it in A$_2$Ir$_2$O$_7$.

\section{Field induced charge-density-wave at pinned wavevector}
The physics discussed in the previous section is essentially of single-particles and can be realized at a rather high temperature. In this section we consider the correlation physics. Without magnetic field a Weyl semi-metal is a stable phase in the presence of correlation, because power counting shows that a short-range interaction is perturbatively irrelevant in the sense of renormalization group. In the following we discuss the correlation-induced \emph{instability} of a Weyl semi-metal in a magnetic field.

A well-known correlated effect of a 3D metal is the CDW instability in a magnetic field\cite{Fukuyama1978783,JPSJ.50.725,PhysRevB.47.8851}. The origin of the instability can be understood easily: in a magnetic field along $z$-direction, the $k_z$ is still a good quantum number and 3D metal forms Landau bands. Because the Landau degeneracy $\propto B$, one expect physics similar to nested Fermi surface occurs at $2k_F$--the CDW instability \emph{along the field direction}.  Transport and magnetometry experiments in metals with a small carrier density have observed evidences of the CDW phases. For example in graphite\cite{PhysRevB.25.5478,JPSJ.68.181,PhysRevLett.54.1182,Iye15011998}, signatures of field-induced CDW such as transport singularity at $T_c$ and non-Ohmic behaviors have been observed. In Bismuth, hysteresis in magnetic torque measurement has been speculated to be associated with formation of CDW\cite{Li25072008,PhysRevB.79.241101}. The CDW phase transition temperature $T_c$ was found $\sim e^{-\frac{B_*}{B}}$ in graphite\cite{PhysRevB.25.5478} ($T_c\sim 1$K for $B\sim25$Tesla), consistent with a BCS type of instability.

Recently, exciting experimental progresses have been made in graphite and Bismuth\cite{behnia,PhysRevLett.103.116802}, where plateaus of transport measurement were found beyond the quantum limit, defined to be the magnetic field strength at which all electrons go into the lowest Landau band. Beyond quantum limit the system should be featureless within single-particle description. The features of transport signal far beyond quantum limit have been speculated to be associated with 3D fractional quantum hall effect. To realize quantum limit with the accessible magnetic field strength, people have to work with metals with small carrier density. However, even for Bismuth where carrier density is already very low, quantum limit is $\sim 9$Tesla.

Because the carrier density vanishes in the clean limit, one important feature of the Weyl semi-metal is that it reaches quantum limit even in a weak magnetic field. Consequently, a Weyl semi-metal is an ideal platform to study 3D correlated quantum hall physics. In the section we study the CDW instability of the Weyl semi-metal. Let us start from the two band model Eq.(\ref{eq:two-orbit}). Surprisingly, unlike a usual 3D metal where CDW occurs along the field direction, we find that in the Weyl semi-metal it is pinned at the wavevector $2\vec k_0$ connecting the two nodes and is \emph{independent} of the field direction.

\begin{figure}
\centerline
{
\includegraphics[width =0.3\textwidth]
{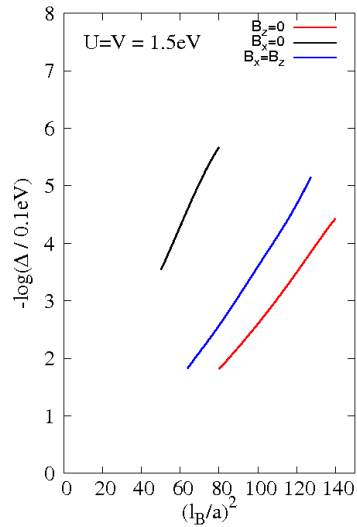}
}
 \caption
 {(color online) The numerical mean-field result of the CDW gap $\Delta$ for the two-band model as a function of magnetic field, expressed in terms of magnetic length $l_B\equiv\frac{\hbar}{eB}$. Three angles of magnetic field: $\vec B\parallel \hat x, \hat z, \mbox{and } \hat x+\hat z$ are shown. CDW is always found to occur at $\vec Q=2\vec k_0$. The exponential dependence $\Delta\sim\Lambda e^{-\frac{B_*}{B}}$ is consistent with a BCS-type instability.
}
\label{fig:CDW} 
\end{figure}

In Fig.\ref{fig:CDW} we present the numerical mean-field calculation of the CDW gap for the two-band model Eq.(\ref{eq:two-orbit}) with $-t_{x} =t_{y} = t_{z} = 0.05 \rm eV$, $m=0.1 \rm eV$, together with the on-site and nearest neighbor repulsions $U=V=1.5$eV. Results are obtained on the lattice system with 65 sites along both $k_{x}$ and $k_{z}$ directions and magnetic field dependent $N_{y} (=(l_{B}/a)^{2}) $ along the $y$ direction ($k_{x}$ and $k_{z}$ are good quantum numbers).
\begin{align}
H_{I} = U\sum_{i, \alpha \neq \beta} n_{i,\alpha}n_{i,\beta} + V \sum_{< i,j >, \alpha ,\beta} n_{i,\alpha}n_{j,\beta} 
\end{align}
where $\alpha, \beta$ are the indices labeling spin, $< i,j >$ represents nearest neighbor. 
In a magnetic field in the $x-z$ plane: $\boldsymbol{B} = B(\sin{\theta},0,\cos{\theta})$, we choose Landau gauge $\boldsymbol{A}(\boldsymbol{r}) = By(-\cos{\theta} , 0, \sin{\theta})$ in which both $k_x,k_z$ are good quantum numbers. After projecting into the Landau bands crossing the Fermi surface, 
\begin{align}
&H'_I=\notag\\
&\sum_{q,\boldsymbol{k},\boldsymbol{k'},i_{1,2,3,4}}U_{\boldsymbol{q},(\boldsymbol{k},i_1,i_2),(\boldsymbol{k'},i_3,i_4)}\gamma^{\dagger}_{\boldsymbol{k+q},i_1}\gamma_{\boldsymbol{k},i_2}(\gamma_{\boldsymbol{k'+q},i_3}^{\dagger}\gamma_{\boldsymbol{k'},i_4})^{\dagger},
\end{align}
 where $\gamma_{\boldsymbol{k},i}$ is the electron in the Landau band labeled by $i$. Diagonalizing the matrix $U_{\boldsymbol{q},(\boldsymbol{k},i_1,i_2),(\boldsymbol{k'},i_3,i_4)}$ for a fixed $\boldsymbol{q}$ gives the most negative eigenvalue $U_{\boldsymbol{q}}$ with its eigenvector $\lambda_{\boldsymbol{k},i_{1},i_{2}}$.  These lead to the mean field Hamiltonian 
\begin{align}
H_{MF} &= \sum_{i} \epsilon_{\boldsymbol{k},i} \gamma_{\boldsymbol{k},i}^{\dag}\gamma_{\boldsymbol{k},i} + \Delta_{\boldsymbol{q}} 
\sum_{i_{1},i_{2}} \lambda_{\boldsymbol{k},i_{1},i_{2}}^{*}  \gamma_{\boldsymbol{k},i_{2}}^{\dag}\gamma_{\boldsymbol{k}+\boldsymbol{q},i_{1}}
+ h.c.
\end{align}
where $\Delta_{\boldsymbol{q}} = U_{\boldsymbol{q}} \langle \sum_{\boldsymbol{k},i_{1},i_{2}} \lambda_{\boldsymbol{k},i_{1},i_{2}} \gamma_{\boldsymbol{k}+\boldsymbol{q},i_{1}}^{\dag} \gamma_{\boldsymbol{k},i_{2}} \rangle$. We then perform a variational mean-field study to find the optimal state to determine the CDW order wavevector and its gap value.

The CDW is found to occur at $\vec Q=(2k_0,0,0)$, independent of the direction of the field. The fact that CDW instability cannot occur along $y$ or $z$ direction can be understood by the following simple physical argument. Let us consider a sample in slab geometry between $z=\pm z_0$. There will be two Fermi arcs at the top and bottom surfaces. As shown in Fig.\ref{fig:semi-classical}, if we view the 3D slab as a 2D sample with a huge unit cell along $z$-direction, it is clear that only when the two Fermi arcs are combined together is a full 2D Fermi surface formed. A CDW can be viewed as layering of the 3D system. If a weak field drives CDW instability along $z$-direction, it would make the top layer isolated from the bottom layer.  If it is true we would end up with a 2D system with a Fermi arc (the concept of Fermi surface is still valid in a weak field), which is not allowed by Luttinger's theorem. Similar argument works for $y$-direction.

Let us now further elaborate this simple physical intuition, and at the same time, show $\vec Q=2\vec k_0$. For the purpose of presentation, let us consider the most striking case: $\boldsymbol{B}=(0,0,B)$ along the $z$-direction. This field still induces the CDW along $x$-direction. For simplicity, let us also assume the Fermi velocity is isotropic around the Weyl nodes. The analytical study of the general case with arbitrary field direction and Fermi velocity anisotropy is discussed in details in appendix. 

In order to understand this CDW pattern, let us consider the low energy Landau bands. At low energy,
\begin{align}
 H=&v_F\psi_R^{\dag}(\vec r)[(-i\vec\partial-\vec k_0+e\vec{A}(\vec r))\cdot(-\sigma_1,\sigma_2,\sigma_3)]\psi_R(\vec r)\notag\\
&+v_F\psi_L^{\dag}(\vec r)[(-i\vec\partial+\vec k_0+e\vec{A}(\vec r))\cdot(\sigma_1,\sigma_2,\sigma_3)]\psi_L(\vec r)
\label{eq:eff}
\end{align}
where $\vec k_0=(k_0,0,0)$. $\psi_{L,R}$ are the electron fields close to $\pm\vec k_0$ in continuum limit.

\begin{figure}
 \includegraphics[width=0.4\textwidth]{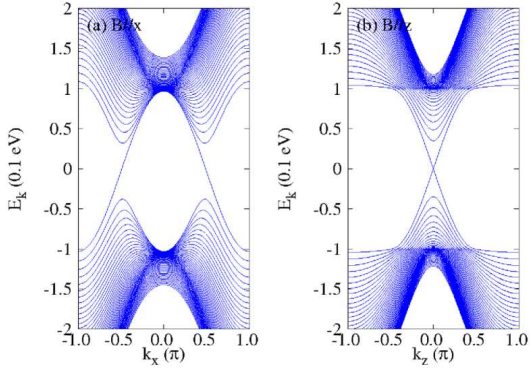}
\caption{(color online) Landau band dispersion for the two-band model with parameters mentioned in the text, when field is along $x$ and $z$ directions respectively. Strength of magnetic field is chosen with $l_B/a=10$}
\label{fig:Landau_band}
\end{figure}

After choosing the Landau gauge $\vec A=(-By,0,0)$, clearly $k_z$ term in Eq.(\ref{eq:eff}) can be viewed as the mass term in the 2D ($k_x,k_y$) Dirac quantum hall problem. Following the well-known result of the energy eigenvalues of a 2D Dirac quantum hall system, we have $E_{L/R,n}=\hbar v_F\mbox{sign}(n)\sqrt{2|n|eB/\hbar+k_z^2}$; i.e., the same energy dispersion for $L$ and $R$ nodes when $n\neq 0$. However, we are focusing on the bands crossing Fermi level, i.e., $n=0$, at which the two branches are counter-propagating: $E_{L/R;0}=\mp\hbar v_F k_z$. The Landau bands for the two-band model Eq.(\ref{eq:two-orbit}) when $\vec B\parallel \hat z, \mbox{ or }\hat x$ are shown in Fig.\ref{fig:Landau_band}.

It is well-known that this Landau level problem can be mapped to a harmonic oscillator. The explicit dependence of the Landau level wavefunction on $y$ for given $k_x,k_z$ is easy to find out: $\xi_{L,0}(y|k_x,k_z)=\big(0,\phi_0(y|k_x+k_0,k_z)\big)$ and $\xi_{R,0}(y|k_x,k_z)=\big(\phi_0(y|k_x-k_0,k_z),0\big)$, where $\phi_0(y|k_x,k_z)\propto e^{-\frac{eB}{2\hbar}(y-y_0(k_x))^2}$ and $y_0(k_x)=\frac{\hbar k_x}{eB}$. 

\begin{figure}
 \includegraphics[width=0.48\textwidth]{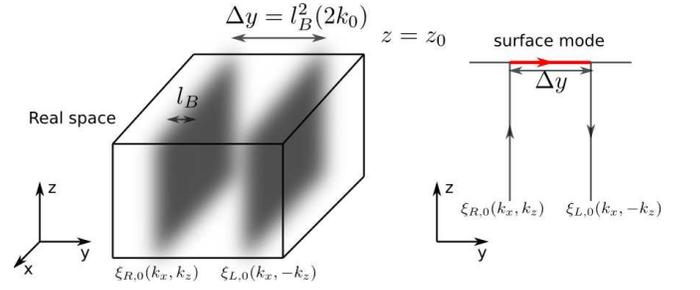}
\caption{(color online)Schematic illustration of the Landau wavefunction reflection problem at surface $z=z_0$.}
\label{fig:reflection}
\end{figure}

Clearly, for the same value of $k_x$, the $L$ mode and $R$ mode are spatially separated by $\Delta y=\frac{\hbar 2k_0}{eB}$. In fact, this spatial displacement is dictating the existence of the surface metallic mode. To see this, one can consider a simple reflection problem of the $z=z_0$ surface (see Fig.\ref{fig:reflection}) while keeping the system translation symmetric along $x,y$ directions. In this setup the $R$ mode is moving along $+z$ direction; after hitting the surface it must be reflected back to the $L$ mode --- the only mode moving along $-z$ direction. Because $k_x$ is a good quantum number, $\xi_{L,0}(y|k_x,-k_z)$ must be connected with $\xi_{R,0}(y|k_x,k_z)$ on the surface! Because they are spatially separated in the bulk, there must be surface modes connecting them. 

In the following we show that this spatial displacement also dictates that the CDW instability can only occur at $\vec Q=2\vec k_0$. In general, the CDW order parameter at momentum $\vec Q$ can be written as $\Delta_{\vec Q}=\sum_{\vec k}f_{\vec k}\gamma_{R,0}^{\dagger}(\vec k+\vec Q)\gamma_{L,0}(\vec k)$, where $f_{\vec k}$ is a profile factor that should be determined energetically. Note that we must have $Q_z=0$ because only the matrix element between the $L$ and $R$ modes at $Q_z=0$ can induce an energy gap. What is $Q_x$?

Because the spatial displacement discussed above, only when $Q_x\sim2k_0$ does the $L,R$ modes overlap spatially. At the mean-field level, the CDW order is coming from  the energy gain by decoupling the repulsive interaction: $U\gamma_{L,0}^{\dagger}\gamma_{L,0}\gamma_{R,0}^{\dagger}\gamma_{R,0}\rightarrow -U\gamma_{L,0}^{\dagger}\gamma_{R,0}\gamma_{R,0}^{\dagger}\gamma_{L,0}$. For a short-range repulsion, this interaction vanishes unless the separation between wavefunctions $\xi_{L/R,0}$ is smaller than the magnetic length $l_B\equiv \sqrt{\frac{\hbar}{eB}}$, indicating $Q_x=2k_0$. 

In appendix we study a generally direction of the magnetic field. In this case we show the spatial separation of $\xi_{L/R,0}$ with momentum difference $\vec Q$ to be $\frac{\hbar}{eB}|\vec Q_{\perp}-2\vec k_{0,\perp}|$, where $\vec Q_{\perp}$ is the component of $\vec Q$ normal to the direction of the field. For the interaction to be effective, $\vec Q_{\perp}=2\vec k_{0,\perp}$. In order to open up an energy gap at Fermi surface,  $\vec Q_{\parallel}=2\vec k_{0,\parallel}$. These two conditions dictate $\vec Q=2\vec k_0$ with a general field direction.

The CDW phase in the two-band model itself is interesting. The modulation of density along $x$-direction can be viewed as spontaneous layering of the 3D homogeneous system into a stacking of 2D layers, where each layer is a period of the CDW. Because the anomalous hall conductivity cannot jump across the phase transition, and because $\nu=2\vec k_0$ as discussed in previous section, it is clear that each CDW layer is exactly $\nu=1$ integer quantum hall layer.

\begin{figure}
 \includegraphics[width=0.25\textwidth]{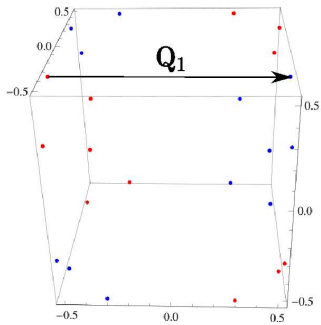}
\caption{(color online) 24 nodes (same as Fig.\ref{fig:BZ}) in the Weyl phase of A$_2$Ir$_2$O$_7$ plotted in a different fashion, it is clear that a wave-vector $\vec Q_1\parallel[1,0,0]$ (and the symmetry related $\vec Q_2\parallel[0,1,0]$,$\vec Q_3\parallel[0,0,1]$, which are not shown) connects 8 pairs of nodes.}
\label{fig:24_nodes}
\end{figure}

Now let us discuss this field induced CDW in the proposed Weyl phase of A$_2$Ir$_2$O$_7$, where 24 nodes are present. Following our result of the simple two-band model, in principle CDW of all the wave-vectors connecting nodes with opposite chirality have instabilities. The true ground state should be determined by energetics. Here we find that there is a particularly likely CDW wave-vector $\vec Q_1\parallel [1,0,0]$ direction, as shown in Fig.\ref{fig:24_nodes}, which connects 8 pairs of nodes with opposite chirality. This means a factor of 8 enhancement of the density of state in the instability. Similarly there are two symmetry related $\vec Q_{2,3}$ along $[0,1,0]$ and $[0,0,1]$ directions respectively. We propose that in the ground state CDW wavevector occurs at these $\vec Q$'s.  If CDW occurs only at one wave-vector, it is a one-dimensional density wave. If CDW of two or three wave-vectors coexist, the ground state would be a two-dimensional or three-dimensional crystal. To tell which phase is realized in A$_2$Ir$_2$O$_7$, one needs higher order terms of the free energy. We leave this question as a topic of future experimental/theoretical investigation.

We remark on the $T_c$ for the CDW phase transition. Dimensional analysis of this BCS-type instability tells us that $T_c\sim \Lambda \exp(-\alpha\frac{\hbar v_F}{u\cdot a}\frac{l_B^2}{a^2})=\Lambda e^{-\frac{B_*}{B}}$, where $\Lambda$ is a cut-off energy scale--typically the bandwidth of the Landau band, $u$ is the effective Hubbard-$U$-type repulsion energy scale, $a$ is the lattice spacing, $l_B\equiv\sqrt{\frac{\hbar}{eB}}$, and $\alpha$ is a dimensionless number. $T_c$ exponentially decays when $B\ll B_*$. In appendix we also consider the effect of a screened Coulomb interaction, and the result can be understood by replacing $u$ here by an energy scale introduced by the screening length. Instead of attempting to compute $\alpha$ and estimate $u$ accurately, let us just compare the $T_c$ of A$_2$Ir$_2$O$_7$ with that of graphite, where experimentally $T_c\sim 1$K for $~25$Tesla field. An estimate based on the known band structures shows that $\frac{\hbar v_F}{a^3}$ is comparable in the two systems. The effective $u$ is hard to estimate because of contributions from the tail of the long-range repulsion, but the naive value $u\sim 5$eV for graphite maybe a factor of $3\sim 4$ larger than that of A$_2$Ir$_2$O$_7$. However the factor of 8 enhancement of density of states in A$_2$Ir$_2$O$_7$ eventually makes its $T_c$ likely to be higher than that of graphite (with the same field strength). Overall our estimate indicates $B_{*}$ for A$_2$Ir$_2$O$_7$ is smaller that of graphite by a factor of $2\sim 3$, making the CDW phase transition proposed here more accessible by various experiment techniques. 

Experimentally, transport measurement directly coupled with the CDW phase transition. Singularities of $\rho_{xx},\rho_{xy}$ at $T_c$ are expected. Non-Ohmic behavior in electric transport is also a signature of a generic CDW\cite{PhysRevB.18.6245}. Moreover, thermodynamic measurements, especially the magnetometry signal (torque measurement), should have singularity at $T_c$. Both experiments can be used to detect the proposed CDW phase transition. Finally we remark on the effect of disorder. A charge disorder is a ``pair-breaking'' defect for the CDW order parameter: the bound state of a particle and a hole. As a result one expects $T_c$ to be reduced by disorder. In graphite the reduction of $T_c$ due to charge impurities has been fitted by the pairing breaking formula of a BCS-type phase transition\cite{Iye15011998}: $\ln(\frac{T_c}{T_{c0}})=\Psi(\frac{1}{2})-\Psi(\frac{1}{2}+\frac{\hbar}{2\pi\tau k_B T_c})$ where $\tau$ is the scattering rate, and $\Psi$ is the digamma function. The reduction of $T_c$ is effective only when $\tau<\frac{\hbar}{k_B T_{c0}}$. In graphite it was found that an impurity density of $2\times 10^{16}$cm$^{-3}$ reduces $T_c$ by $\sim 30\%$ at $B=20$Tesla\cite{Iye15011998}. This provides a rough estimate of the required quality of the sample to observed the proposed CDW phase transition in A$_2$Ir$_2$O$_7$.

\section{Concluding remark}
In this paper we study the responses of a general Weyl semi-metal in a magnetic field. Two novel effects are predicted: a pressure-induced anomalous hall effect, and a field-induced CDW at pinned wavevector, both are intrinsic signatures of the Weyl semi-metal phase. We also applied these general results to the proposed Weyl phase in pyrochlore Iridate.

The pressure-induced anomalous hall effect is a large effect and stable towards temperature and disorder. Our model calculation of the proposed Weyl phase in A$_2$Ir$_2$O$_7$ shows $1\%$ change of the band-structure due to a pressure along [1,1,1] gives rise to in-plane anomalous hall conductivity $\sigma_{AH} \sim 4\Omega^{-1}/$cm. Transport experiments in the near future in these systems can be used to verify/falsify the proposed Weyl phase. \emph{The predicted $P$-linear dependence of anomalous hall conductivity, together with the zero carrier density in the absence of pressure, is a unique property of the Weyl semi-metal phase.} Such a \emph{tunable} anomalous hall effect (from zero to large) may be useful for applications in the future.

We estimate that the $T_c$ of the CDW phase transition is higher than that of graphite in the same field strength. However, to experimentally observe the correlated CDW phase, one still needs a clean sample in a strong magnetic field. Another possible complication specific for the pyrochlore Iridates compounds, which we did not discuss here, is the possible field-induced magnetic structure phase transition. For example, the LDA calculation\cite{PhysRevB.83.205101} estimates that the energy difference of different magnetic ordering patterns is $\sim 3$meV per unit cell.  It is possible that a high magnetic field $\sim 20$Tesla causes the competing magnetic ordered phases to come into play. Our proposed CDW phase transition, as an instability, should be realized at least in the field strength before the possible magnetic order transition. From both experimental and theoretical point of views, these alternative possibilities of pyrochlore Iridates in a high magnetic field are also very interesting and deserve further investigation.

YR and KYY are supported by the start-up fund at Boston College. KYY is also partially supported by DOE-DE-SC0002554, YML is supported by DOE grant DE-FG02-99ER45747. We appreciate helpful discussion with Ziqiang Wang and comments from Ashvin Vishwanath.

\begin{appendix}

\section{Analytical mean field calculation for CDW instability, the general case}

To focus on the low-energy physics around the two Dirac cones
located at $\pm\tk_0=(\pm k_0,0,0)$ in the 3-D 1st BZ, we introduce
the field $\psi({\bf r})\sim a^{-\frac32}c_{\bf r}$ in the continuum
limit where $c_{\bf r}$ is the Fermion annihilation operator in the
lattice model with $a$ being the lattice constant. The Dirac Fermion
at $\tk_0$ coupled with $U(1)$ electromagnetic gauge field $\tA$
through the minimal coupling is described by
\begin{equation}
H_0=v_F\int\text{d}^3{\bf r}\psi^\dagger({\bf
r})\big(-\imth\hbar\mathbf\nabla-\hbar\tk_0+e\tA(\bf
r)\big)\cdot\mathbf\sigma~\psi(\bf r) 
\end{equation}
where $e=|e|$ and the electron charge is $-e$. Fermi velocity
$v_F\sim ta/\hbar$ where $t$ is the hopping energy in the lattice
model. Without loss of generality, we consider a constant magnetic
field $\mathbf{B}=(B_x,0,B_z)\equiv B(\sin\theta,0,\cos\theta)$
under Landau gauge: $\tA({\bf
r})=(-B_zy,0,B_xy)=By(-\cos\theta,0,\sin\theta)$. This problem in
real space
\begin{eqnarray}\label{H0}
H_0({\bf r})=v_F\big(-\imth\hbar\mathbf\nabla-\hbar\tk_0+e\tA(\bf
r)\big)\cdot\mathbf\sigma
\end{eqnarray}
can be exactly solved since
\begin{eqnarray}
&\notag\big(H_0({\bf r})\big)^2=(\hbar
v_F)^2\big[(-\imth\partial_x-k_0-\frac{eB_zy}\hbar)^2+\\
&(-\imth\partial_z+\frac{eB_xy}\hbar)^2\big]-(v_F\hbar\partial_y)^2+v_F^2e\hbar~{\bf
B}\cdot\mathbf\sigma
\end{eqnarray}
is nothing but a Harmonic oscillator diagonalized by ladder
operators
$b=\sqrt{\frac{eB}{2\hbar}}(y-y_0)+\sqrt{\frac{\hbar}{2eB}}\partial_y$
and $b^\dagger$. In the Landau gauge the Hamiltonian (\ref{H0}) is
manifestly invariant under translations along $\hat x$ and $\hat z$
directions. Therefore the eigenstates with energy
$E_n(k_x,k_z)=\hbar
v_F~\text{sign}(n)\sqrt{2|n|eB/\hbar+[k_z\cos\theta+(k_x-k_0)\sin\theta]^2}$
are labeled by momenta $k_{x,z}$ and Landau level index
$n=0,\pm1,\pm2,\cdots$. It's convenient to introduce the magnetic
length $l_B\equiv\sqrt{\hbar/eB}$ and the Hamiltonian (\ref{H0}) can
be simplified as
\begin{eqnarray}
&H_0={\hbar
v_F}e^{-\imth\frac\theta2\sigma_y}\begin{bmatrix}k_{||}&-\sqrt2~b/l_B\\-\sqrt2~b^\dagger/l_B&-k_{||}\end{bmatrix}e^{\imth\frac\theta2\sigma_y}\notag
\end{eqnarray}
where we define $k_{||}\equiv k_z\cos\theta+(k_x-k_0)\sin\theta$.
Apparently the wavefunctions for Landau levels $n\neq0$ are
\begin{eqnarray}
&\xi_n(y|k_x,k_z)=e^{-\imth\frac\theta2\sigma_y}\cdot\notag\\
&\begin{bmatrix}\frac{\sqrt{2|n|}}{l_B}\phi_{|n|-1}(y|k_x-k_0,k_z)\\
\big(k_{||}-\text{sign}(n)\sqrt{k^2_{||}+2|n|/l_B^2}\big)\phi_{|n|}(y|k_x-k_0,k_z)\end{bmatrix}\notag
\end{eqnarray}

Especially for $n=0$ Landau level the energy and wavefunction of
eigenstates are
\begin{eqnarray}
\notag&E_0(k_x,k_z)=-v_F\hbar\big[k_z\cos\theta+(k_x-k_0)\sin\theta\big],\\
&\xi_0(y|k_x,k_z)=(-\sin\frac\theta2,\cos\frac\theta2)^T\phi_0(y|k_x-k_0,k_z).
\end{eqnarray}
where
\begin{eqnarray}
&\phi_n(y|k_x-k_0,k_z)\equiv\\
&\notag(\frac{eB}{\pi\hbar})^{\frac14}\frac1{\sqrt{n!}}\Big(\sqrt{\frac{eB}{2\hbar}}(y-y_0)-\sqrt{\frac{\hbar}{2eB}}\partial_y\Big)^ne^{-\frac{eB}{2\hbar}(y-y_0)^2},\\
&
y_0\equiv\frac{\hbar}{eB}\big[(k_x-k_0)\cos\theta-k_z\sin\theta\big].\notag
\end{eqnarray}
Notice that the energy $E_n(k_x,k_z)$ only disperses along the
direction of the magnetic field, \ie $\partial E_n(k_x,k_z)/\partial
k_{\perp}=0$ with $k_\perp\equiv\cos\theta(k_x-k_0)-\sin\theta k_z$.
This is the Landau degeneracy of energy levels under a magnetic
field.\\

Now consider two branches Dirac Fermions (left-moving branch
$\psi_L$ and right-moving $\psi_R$) at $\pm\tk_0$ with opposite
chirality (\ie the {sign} of $v_xv_yv_z$). The electron field is
expressed as
\begin{eqnarray}
&\psi({\bf r})\sim\sum_{\tk\simeq-\tk_0}e^{\imth\tk\cdot{\bf
r}}\psi_{L,\tk}+\sum_{\tk\simeq+\tk_0}e^{\imth\tk\cdot{\bf
r}}\psi_{R,\tk}\notag\\
&\notag=\frac1\Omega\sum_{k_z}e^{\imth k_z
z}\sum_n\Big(\sum_{k_x\simeq-k_0}e^{\imth
k_xx}\xi_n^L(y|k_x,k_z)\gamma_n^L(k_x,k_z)\\
&\notag+\sum_{k_x\simeq+k_0}e^{\imth
k_xx}\xi_n^R(y|k_x,k_z)\gamma_n^R(k_x,k_z)\Big)
\end{eqnarray}
where $\gamma_n^{L,R}(k_x,k_z)$ are the annihilation operators for
the eigenmodes in the $n$-th Landau bands with momenta $k_x,k_z$.
Under a magnetic field both branches have a huge Landau degeneracy
which could easily cause Fermi surface nesting and hence the
charge-density-wave (CDW) instability with a nesting vector
$\pm2\tk_0$. In other words, an infinitely small interaction (\eg
on-site Hubbard-type interaction) might drive the system into a CDW
phase. For simplicity we consider the following on-site Hubbard-type
repulsive interaction
\begin{equation}\label{int}
V_H=\frac12\sum_{\alpha,\beta}\int\text{d}^3{\bf
r}~u_{\alpha\beta}\big[\psi^\dagger_\alpha({\bf r})\psi_\alpha({\bf
r})\big]\big[\psi^\dagger_\beta({\bf r})\psi_\beta({\bf r})\big]
\end{equation}
where $\alpha,\beta$ are band/spin indices. In the continuum model
$u_{\alpha\beta}\sim U_{\alpha\beta}a^3$ where $U_{\alpha\beta}$ are
the on-site Hubbard repulsion energy in the lattice model.
Meanwhile, to concentrate on the low-energy physics we restrict our
study to the $n=0$ Landau band $E_0^{L,R}(k_x,k_z)=\pm
v_F\hbar\big[k_z\cos\theta+(k_x\pm k_0)\sin\theta\big]$ with
wavefunctions $\xi^{L,R}(y|k_x,k_z)$. But notice in different models
the two Dirac cones can be on the same orbits or not: in other words
they could have different band indices (at least in the low-energy
limit).

In general by projecting the interaction (\ref{int}) into the $n=0$
Landau level we have the following terms that contribute to the
$\pm2\tk_0$ scattering
\begin{eqnarray}
&\label{cdw channel}
V_{CDW}=-\frac1\Omega\sum_{k_{x,z},k^\prime_{k,z}}V_{k_{x,z}|k^\prime_{x,z}}{\gamma^L_0}^\dagger(k_x-k_0,k_z)  \nonumber \\
&\gamma^R(k_x+k_0,k_z){\gamma^R_0}^\dagger(k_x^\prime+k_0,k_z^\prime)\gamma^L(k^\prime_x-k_0,k^\prime_z)
\end{eqnarray}
where $\Omega$ is the sample size in $\hat x$-$o$-$\hat z$ plane.
The Fourier transformation is defined as
\begin{eqnarray}
\notag\psi({\bf r})=\frac1{\sqrt\Omega}\sum_\tk e^{\imth\tk\cdot{\bf
r}}\psi_\tk
\end{eqnarray}

The CDW order parameter is
\begin{eqnarray}\label{order_param}
&\Delta_{k_{x,z}}\equiv\frac1\Omega\sum_{k^\prime_x,k^\prime_z}V_{k_{x,z}|k^\prime_{x,z}}\cdot\\
&\notag\langle{\gamma^R_0}^\dagger(k^\prime_x+k_0,k^\prime_z)\gamma^L(k^\prime_x-k_0,k^\prime_z)\rangle
\end{eqnarray}
Notice that here $k_{x,z}$ and $k^\prime_{x,z}$ are all small
momenta upper-bounded by an ultraviolet cutoff $\Lambda<2k_0$.  The
bare coupling constants $V_{k_{x,z}|k^\prime_{x,z}}$ are calculated
by
\begin{eqnarray}\label{Vcdw}
& V_{k_{x,z}|k^\prime_{x,z}}\equiv\sum_{\alpha\beta}u_{\alpha\beta}\int\text{d}y\cdot\\
&\notag\Big\{{\xi_\alpha^L}^\ast(y|k_x-k_0,k_z)\xi_\alpha^L(y|k_x^\prime-k_0,k_z^\prime)\cdot\\
&\notag{\xi_\beta^R}^\ast(y|k^\prime_x+k_0,k^\prime_z)\xi_\beta^R(y|k_x+k_0,k_z)(1-\delta_{k_x,k_x^\prime}\delta_{k_z,k_z^\prime})\\
&\notag-{\xi_\alpha^L}^\ast(y|k_x-k_0,k_z)\xi_\alpha^R(y|k_x+k_0,k_z)\cdot\\
&\notag{\xi_\beta^R}^\ast(y|k_x^\prime+k_0,k_z)\xi_\beta^L(y|k_x^\prime-k_0,k_z^\prime)\Big\}
\end{eqnarray}
In the 1st term the $\textbf{q}=\tk-\tk^\prime=0$ component is
canceled by contributions  uniform positive charge background (to
keep the total charge neutrality).

\subsection{Effects of Fermi velocity anisotropy}

Now let's consider a more general case, \ie a Dirac cone at $\tk_0$ with
anisotropic Fermi velocities $v_m,~m=x,y,z$. In this case we define
$v_F=v_y$ and rescale the coordinates by $\tilde
x=\frac{v_F}{v_x}x$, $\tilde y=y$ and $\tilde z=\frac{v_F}{v_z}z$
and the Hamiltonian
\begin{eqnarray}
H_0=\sum_{m=x,y,z}v_m\big(-\imth\hbar\mathbf\nabla-\hbar\tk_0+e\tA({\bf
r})\big)_m\cdot\mathbf\sigma_m\notag
\end{eqnarray}
has the form of (\ref{H0}) in the new coordinate system. But in the
Landau gauge, the vector potential in the rescaled coordinate system
becomes $\tilde\tA(\textbf{r}^\prime)=\tilde
y(-\frac{v_x}{v_y}B_z,0,\frac{v_z}{v_y}B_x)$, \ie the effective
magnetic field in rescaled coordinate system is
$\tilde{\textbf{B}}=(\frac{v_z}{v_y}B_x,0,\frac{v_x}{v_y}B_z)\equiv
\tilde B(\sin\tilde\theta,0,\cos\tilde\theta)$. Meanwhile the
momentum transforms as
$\tilde\tk=(\frac{v_x}{v_y}k_x,k_y,\frac{v_z}{v_y}k_z)$. On the
other hand, to keep the action invariant, we need to rescale the
filed $\tilde\psi({\bf
r}^\prime)=\sqrt{|\frac{v_xv_z}{v_y^2}|}\psi({\bf r})$ and the
rescaled interaction coupling constants become $\tilde
u=u\frac{v_y^2}{|v_xv_z|}$. And all conclusions discussed earlier
can be adopted by replacing $\{\textbf{r},\tk,\tk_0,B,\theta,u\}$
with
$\{\tilde{\textbf{r}},\tilde\tk,\tilde{\tk_0},\tilde{B},\tilde\theta,\tilde
u\}$. In the following calculations we shall ignore the
$\tilde{}$~notation and assume all quantities are rescaled ones
unless specifically mentioned.\\

\subsection{Four-band model}
For reason that will become clear soon, it is useful to study another tight-binding realizing two Weyl nodes:
\begin{align}
 H=&v_F(\sin \vec k\cdot \vec \sigma\tau^3-k_0\sigma^1\tau^0)\notag\\
&-m[3-(\cos k_x+\cos k_y+\cos k_z)]\sigma^0\tau^1
\end{align}
where $\sigma^0,\tau^0$ both are the identity 2 by 2 matrix.

In this case the two Dirac cones come from different orbits and
their $n=0$ Landau bands have no overlap with each other in the
low-energy limit. We focus on the simplest case in which they have
opposite Fermi velocities $\pm v_{x,y,z}$. The non-interacting
Hamiltonian for 4-component Fermion field $\psi({\bf
r})=(\psi_R({\bf r}),\psi_L({\bf r}))$ are
\begin{eqnarray}
H_{4b}({\bf r})=v_F\Big[\big(-\imth\hbar\mathbf\nabla+e\tA({\bf
r})\big)\tau^3-\hbar\tk_0\tau^0\Big]\cdot\mathbf\sigma
\end{eqnarray}
After considering the anisotropy the Fermi velocity we find that the
two Dirac cones have the same rescaled magnetic field \ie
$\tilde\theta_L=\tilde\theta_R\equiv\theta$. It's easy to find that
\begin{eqnarray}
&\notag\cos\theta=\text{sign}(v_xv_yv_z)\frac{B_z/v_z}{\sqrt{(B_x/v_x)^2+(B_z/v_z)^2}},\\
&\notag\sin\theta=\text{sign}(v_xv_yv_z)\frac{B_x/v_x}{\sqrt{(B_x/v_x)^2+(B_z/v_z)^2}}.
\end{eqnarray}
and the rescaled magnitude of magnetic field is
$B=\sqrt{(B_xv_z)^2+(B_zv_x)^2}/|v_y|$. The wavefunctions of $n=0$
Landau level eigenstates are
\begin{eqnarray}
&\notag\xi^{L}(y|k_x,k_z)\equiv(0,0,-\sin\frac\theta2,\cos\frac\theta2)^T\phi_0(y|k_x+
k_0,k_z),\\
&\notag\xi^{R}(y|k_x,k_z)\equiv(-\sin\frac\theta2,\cos\frac\theta2,0,0)^T\phi_0(y|k_x-
k_0,k_z).
\end{eqnarray}
where wavefunction $\phi_0(y|k_x,k_z)$ is a Gaussian wavepacket
centered at
\begin{eqnarray}
y_0=l_B^2(\cos\theta\tilde k_x-\sin\theta\tilde k_z)=\frac\hbar
e\cdot\frac{\frac{B_z}{v_z^2}k_x-\frac{B_x}{v_x^2}k_z}{(\frac{B_z}{v_z})^2+(\frac{B_x}{v_x})^2}
\end{eqnarray}

Therefore the 2nd term in (\ref{Vcdw}) vanishes and there are CDW
instabilities at momentum ${\bf Q}=2\tk_0$. In the simplest case
when $u_{\alpha\beta}\equiv u$ we have
\begin{eqnarray}\label{bare coupling}
V_{k_{x,z}|k^\prime_{x,z}}=\frac
u{\sqrt{2\pi}l_B}e^{-\frac12l_B^2\big((\tilde k_x-\tilde
k_x^\prime)\cos\theta-(\tilde k_z-\tilde
k_z^\prime)\sin\theta\big)^2}
\end{eqnarray}

In the following we study the mean-field theory of CDW in such a
four-band system. First we consider the case with isotropic Fermi
velocities:
\begin{eqnarray}\label{mf_ham}
H_{MF}=\sum_{|k_{x,z}|<\Lambda}\begin{pmatrix}\gamma^L_k\\
\gamma^R_{k}\end{pmatrix}^\dagger\begin{bmatrix}\hbar v_Fk_{||}&-\Delta_k\\-\Delta_k^\ast&-\hbar v_Fk_{||}\end{bmatrix}\begin{pmatrix}\gamma^L_k\\
\gamma^R_{k}\end{pmatrix}
\end{eqnarray}
where we denote $\gamma^{L,R}_k\equiv\gamma^{L,R}_0(k_x\mp k_0,k_z)$
and $k_{||}\equiv k_x\sin\theta+k_z\cos\theta$. The self-consistent
conditions for the order parameters are
\begin{eqnarray}\label{gap_eq}
\Delta_k=\frac1\Omega\sum_{|k^\prime_{x,z}|<\Lambda}V_{k_{x,z}|k^\prime_{x,z}}\frac{\Delta_{k^\prime}}{2E_{k^\prime}}\big(1-2f(E_{k^\prime})\big)
\end{eqnarray}
where $E_k=\sqrt{(\hbar v_Fk_{||})^2+|\Delta_k|^2}$ are the
eigenvalues of mean-field Hamiltonian (\ref{mf_ham}) and
$f(\epsilon)=\big[1+\exp(\beta\epsilon)\big]^{-1}$ is the Fermi
distribution function. Choosing a new coordinate system $k_{||}$ and
$k_\perp=k_x\cos\theta-k_z\sin\theta$, we can see that (\ref{bare
coupling}) would decay exponentially with
$(|k_\perp-k^\prime_\perp|l_B)^2$. The magnetic length has the order
of magnitude $\sim\frac{257}{\sqrt{B(T)}}{\AA}$, which is about a
hundred times larger than the lattice constants. As a result we can
safely ignore the momentum dependence of order parameter, \ie
$\Delta_k\approx\Delta$ to a very good approximation as long as the
ultraviolet cutoff $\Lambda\gg1/l_B$. Plugging (\ref{mf_ham}) into
self-consistent equation (\ref{gap_eq}) and performing the $k_\perp$
integration we have
\begin{equation}
2=\frac{u}{(2\pi
l_B)^2}\int_{|k_{||}|<\Lambda}\text{d}k_{||}\frac{\tanh(\beta\sqrt{|\Delta|^2+(\hbar
v_Fk_{||})^2})}{\sqrt{|\Delta|^2+(\hbar v_Fk_{||})^2}}
\end{equation}
At zero temperature the order parameter is determined by
\begin{eqnarray}
\log\big[\sqrt{1+(\frac{\hbar v_F\Lambda}{\Delta})^2}+\frac{\hbar
v_F\Lambda}{\Delta}\big]=(2\pi l_B)^2\hbar v_F/u\gg1\notag
\end{eqnarray}
where we choose a proper gauge so that $\Delta_k=|\Delta_k|$. This
indicate the critical temperature or the CDW energy gap is
\begin{eqnarray}
\notag\Delta\sim k_BT_c\sim2\hbar v_F\Lambda e^{-\frac{(2\pi
l_B)^2\hbar v_F}{u}}\sim 2t~e^{-\#\frac tU(\frac{2\pi l_B}a)^2}
\end{eqnarray}
where $t$ and $U$ are the hopping and interaction energy scale in
the lattice model. $\#$ is a constant of order 1.

Considering the anisotropy of Fermi velocities, we get the following
estimation of $T_c$ or $\Delta$:
\begin{eqnarray}
\Delta\sim
k_BT_c\sim2t~e^{-\frac{(2\pi\hbar)^2}{eu}\big[(\frac{B_x}{v_x})^2+(\frac{B_z}{v_z})^2\big]^{-1/2}}
\end{eqnarray}
Apparently aligning the magnetic field along the direction with a
smaller Fermi velocity would result in a higher critical temperature
$T_c$.\\

\subsection{Two-band model}

Now let us come back to the two-band model in the main text Eq.(\ref{eq:two-orbit}). In this case the two Dirac components can be viewed coming from the spin index.

In the simplest case where the two Dirac cone has opposite Fermi
velocities (therefore $\theta^\prime_L=\theta^\prime_R\equiv\theta$
for rescaled magnetic field $\textbf{B}_{L,R}^\prime$ as in the
four-band case), we have
\begin{equation}
\xi^{L/R}(y|k_x,k_z)\equiv(-\sin\frac{\theta}2,\cos\frac{\theta}2)^T\phi_0(y|k_x\pm
k_0,k_z)
\end{equation}
Clearly since
$\xi^L(y|k_x-k_0,k_z)=\xi^R(y|k_x+k_0,k_z)=(-\sin\frac{\theta}2,\cos\frac{\theta}2)^T\phi_0(y|k_x,k_z)$
and the two terms in (\ref{Vcdw}) exactly cancel each other. So we
don't have CDW instability in this case.

Let's consider a little more complicated case, which is numerically
studied in a two-band tight-binding model where the two Dirac cones
share the same $v_{y,z}$ but opposite $v_x$. In this case after
rescaling we have $\tilde{\textbf{B}}_L=(\tilde B_x,0,\tilde B_z)$
and $\tilde{\textbf{B}}_R=(\tilde B_x,0,-\tilde B_z)$ and thus
$\tilde\theta_L=\theta,~\tilde\theta_R=\pi-\theta$. The
wavefunctions for the two branches read
\begin{eqnarray}
&\notag\xi^{L}(y|k_x,k_z)\equiv(-\sin\frac{\theta}2,\cos\frac{\theta}2)^T\phi_0(y|k_x-k_0,k_z),\\
&\notag
\xi^{R}(y|k_x,k_z)\equiv(-\cos\frac{\theta}2,\sin\frac{\theta}2)^T\phi_0(y|k_x+
k_0,k_z).
\end{eqnarray}
As a result the 2nd term in (\ref{Vcdw}) contributes
$\sim(2\sin\frac\theta2\cos\frac\theta2)^2=\sin^2\theta$ and in the
simplest case with $u_{\alpha\beta}\equiv u$ (corresponding to only having on-site $U$) we have
\begin{equation}\label{bare coupling_2band}
V_{k_{x,z}|k^\prime_{x,z}}=\frac
{u\cos^2\theta}{\sqrt{2\pi}l_B}e^{-\frac12l_B^2\big((\tilde
k_x-\tilde k_x^\prime)\cos\theta-(\tilde k_z-\tilde
k_z^\prime)\sin\theta\big)^2}
\end{equation}
Following derivations are completely similar with the four-band case
and in the end we have
\begin{eqnarray}
\notag\Delta\sim k_BT_c\sim2t~e^{-\frac{(2\pi l_B)^2\hbar
v_F}{u\cos^2\theta}}\sim2t~e^{-\frac{(2\pi\hbar
v_z)^2}{euB_z^2}\sqrt{(\frac{B_x}{v_x})^2+(\frac{B_z}{v_z})^2}}
\end{eqnarray}
Clearly when magnetic field is along $\hat x$-axis \ie $B_z=0$ there
are no CDW orders even at zero temperature. But this is because two band model with only on-site $U$ ( $u_{\alpha\beta}\equiv u$ ) is sick in this limit. After turning on a nearest neighbor repulsion $V$, as shown in the main text, or considering the four-band model as computed in previous section, field along any direction generate CDW gap.

\subsection{Effects of long-range Coulomb interaction}

To compare with on-site Hubbard-type repulsive interaction discussed earlier, here let's consider a long-range Coulomb interaction
\begin{eqnarray}
V_C=\int\text{d}^3\textbf{r}_1~\text{d}^3\textbf{r}_2V_c(|{\bf r}_1-{\bf r}_2|)\big[\psi^\dagger_\alpha({\bf
r}_1)\psi_\alpha({\bf r}_1)\big]\big[\psi^\dagger_\beta({\bf
r}_2)\psi_\beta({\bf r}_2)\big]\notag
\end{eqnarray}
where $\epsilon$ is the dielectric constant in the material. Without loss of generality, we start from a screened Coulomb interaction
\begin{eqnarray}
V_c(|\textbf{r}_1-\textbf{r}_2|)=\frac{e^2\exp(-k_s|\textbf{r}_1-\textbf{r}_2|)}{4\pi\epsilon|\textbf{r}_1-\textbf{r}_2|}\notag
\end{eqnarray}
where $1/k_s$ is the typical screening length. The bare long-range Coulomb interaction corresponds to the case in which $k_s=0$. 

Notice that \begin{eqnarray}
&\notag\int\text{d}x\int\text{d}z\frac{e^{\imth(k_xx+k_zz)}e^{-k_s\sqrt{x^2+y^2+z^2}}}{\sqrt{x^2+y^2+z^2}}\\
&\notag=\frac{2\pi}{\sqrt{k^2+k_s^2}}e^{-y\sqrt{k^2+k_s^2}},~~~k\equiv\sqrt{k_x^2+k_z^2}.\notag
\end{eqnarray}
so the coupling constants in (\ref{cdw channel}) for Coulomb interaction are
\begin{eqnarray}\label{Vcdw_coulomb}
& V_{k_{x,z}|k^\prime_{x,z}}=\frac{e^2}{2\epsilon}\sum_{\alpha\beta}\int\text{d}y_1\int{\text{d}}y_2\cdot\\
&\notag\Big\{\frac{e^{-|y_1-y_2|\sqrt{\delta k^2+k_s^2}}}{\sqrt{\delta k^2+k_s^2}}{\xi_\alpha^L}^\ast(y_1|k_x-k_0,k_z)\xi_\alpha^L(y_1|k_x^\prime-k_0,k_z^\prime)\cdot\\
&\notag{\xi_\beta^R}^\ast(y_2|k^\prime_x+k_0,k^\prime_z)\xi_\beta^R(y_2|k_x+k_0,k_z)(1-\delta_{k_x,k_x^\prime}\delta_{k_z,k_z^\prime})\\
&\notag-~\frac{e^{-\sqrt{4k_0^2+k_s^2}|y_1-y_2|}}{\sqrt{4k_0^2+k_s^2}}{\xi_\alpha^L}^\ast(y_1|k_x-k_0,k_z)\xi_\alpha^R(y_1|k_x+k_0,k_z)\cdot\\
&\notag{\xi_\beta^R}^\ast(y_2|k_x^\prime+k_0,k_z)\xi_\beta^L(y_2|k_x^\prime-k_0,k_z^\prime)\Big\}
\end{eqnarray}
where we denote $\delta
k\equiv\sqrt{(k_x-k_x^\prime)^2+(k_z-k^\prime_z)^2}$. Again total
charge neutrality removes the $\textbf{q}=\tk-\tk^\prime=0$
component in the 1st term. Like discussed earlier in the four-band
case the 2nd term vanishes, while both terms contribute in the
two-band case with opposite $v_x$ for two Dirac cones.

By defining
$k_\perp^2l^2=\frac{|v_xv_z|}{|v_y|}\frac\hbar
e\frac{(\frac{B_z}{v_z^2}k_x-\frac{B_x}{v_x^2}
k_z)^2}{\big[(\frac{B_z}{v_z})^2+(\frac{B_x}{v_x})^2\big]^{3/2}}$ we
can write the coupling constants in this case as:

(\Rmnum{1}) For a four-band model with opposite Fermi velocities
$\pm v_{x,y,z}$ at two Dirac cones
\begin{eqnarray}
&\notag
V_{k_{x,z}|k^\prime_{x,z}}^{4b}=\frac{e^2\text{Erfc}(\frac{l_B\sqrt{\delta
k^2+k_s^2}}{\sqrt2})}{2\epsilon\sqrt{\delta
k^2+k_s^2}}e^{\frac{l_B^2(\delta k^2+k_s^2)-(l\delta
k_\perp)^2}{2}}\end{eqnarray}

(\Rmnum{2}) For a two-band model with opposite $v_x$ but the same
$v_{y,z}$ at two Dirac cones
\begin{eqnarray}&V_{k_{x,z}|k^\prime_{x,z}}^{2b}=V_{k_{x,z}|k^\prime_{x,z}}^{4b}-\frac{e^2\sin^2\theta}{2\epsilon\sqrt{
4k_0^2+k_s^2}}e^{\frac{l_B^2(4k_0^2+k_s^2)-(l\delta
k_\perp)^2}2}\cdot\notag\\
&\notag\text{Erfc}(l_B\sqrt{2k_0^2+k_s^2/2})\approx V_{k_{x,z}|k^\prime_{x,z}}^{4b}\notag
\end{eqnarray}
where
$\text{Erfc}(z)\equiv\frac{2}{\sqrt\pi}\int^{+\infty}_ze^{-t^2}\text{d}t$
is the complementary error function. The 2nd term is ignored since we assume $k_0l_B\gg1$. Notice that the singularity in $V_{k_{x,z}|k^\prime_{x,z}}^{4b}$
when $\delta k_{x,z}=0$ is removed by the screening of bare Coulomb
interaction.

In this case of screened Coulomb interaction, the self-consistent
gap equation (\ref{gap_eq}) turns out to be
\begin{eqnarray}\label{gap_eq_scr}
&\frac{4\epsilon\Delta_k}{e^2}=\int_{|k^\prime|<\Lambda}\frac{\text{d}k^\prime_\perp\text{d}k^\prime_{||}}{(2\pi)^2}\frac{\Delta_{k^\prime}}{\sqrt{|\Delta_{k^\prime}|^2+(\hbar
v_Fk^\prime_{||})^2}}\cdot\\
&\notag\frac{e^{\frac{l_B^2(|k_{||}-k^\prime_{||}|^2+k_s^2)}2}
\text{Erfc}(\frac{l_B\sqrt{|k-k^\prime|^2+k_s^2}}{\sqrt2})}
{\sqrt{|k-k^\prime|^2+k_s^2}}{\tanh(\frac{\sqrt{|\Delta_{k^\prime}|^2+(\hbar
v_Fk^\prime_{||})^2}}{2k_BT})}.
\end{eqnarray}
Notice that $|\Delta_k|\rightarrow0$ as the temperature approaches $T_c$. Although the above gap equation cannot be solved analytically, using the asymptotic behavior of complementary error function 
\begin{eqnarray}
\text{Erfc}(x)\sim\frac{e^{-x^2}}{x\sqrt\pi},~~~x\gg1.\label{asymp_erfc}
\end{eqnarray}
we can figure out the asymptotic behavior of critical temperature as
\begin{eqnarray}
T_c\sim\hbar v_F\Lambda e^{-(c_1+c_2k_s^2l_B^2)\epsilon\hbar v_F/e^2}
\end{eqnarray}
where $c_{1,2}$ are constants of order 1. In the case of bare (unscreened) Coulomb interaction, the CDW critical temperature behaves as 
\begin{eqnarray} 
T_c\sim t~e^{-\#\epsilon\hbar v_F/e^2}
\end{eqnarray}
for both two-band and four-band models, where $t$ denotes the band width. This result can be easily understood by dimensional analysis. An unscreened Coulomb interaction does not give new length scales, and the dimensionless exponent in a BCS-type formula must be $\sim \frac{\epsilon \hbar v_F}{e^2}$. This is in accordance with previous studies in a different formulation\cite{Fukuyama1978783} and suggests that $T_c$ won't change with the magnetic field in the case of bare long-range Coulomb interaction, which is inconsistent with the experimental observation in graphite system.

\end{appendix}

\bibliographystyle{apsrev}
\bibliography{/home/ranying/downloads/reference/simplifiedying}

\end{document}